\documentclass[11 pt]{article}

\usepackage[T1]{fontenc}
\usepackage{amsfonts,mathtools, amsmath,amsthm,amssymb}
\usepackage{graphicx,booktabs,xcolor,float,rotating}
\usepackage[top = 2cm , bottom= 2cm , left= 2cm , right= 2cm]{geometry}
\usepackage[colorlinks,citecolor=gr,linkcolor=gr,urlcolor=gr]{hyperref}

\definecolor{prp}{HTML}{b16286}
\definecolor{bl}{HTML}{4E09B2}
\definecolor{gr}{HTML}{61635F}
\usepackage{algorithm2e}
\usepackage{setspace}
\usepackage[shortlabels]{enumitem}
\usepackage{authblk}

\usepackage{mdframed}
\mdfdefinestyle{st}{%
    linecolor=bl,
   linewidth=2pt,
   leftmargin=-20,
   rightmargin=-20,
        }

\usepackage{titlesec}
\titleformat{\section}
  {\centering\bfseries}
  {\thesection}
  {.5em}
  {\MakeUppercase}

\titleformat{\subsection}
  {\normalfont\bfseries\itshape}{\thesubsection}{1em}{}

\usepackage[noabbrev,capitalise]{cleveref}

\newtheorem{theorem}{Theorem}[section]

\newtheorem{lemma}[theorem]{Lemma}

\usepackage[sort]{natbib}
\usepackage{dsfont}

\newcommand{\btheta}{\boldsymbol{\vartheta}}

\newcommand{\bi}{\boldsymbol{i}}
\newcommand{\by}{\boldsymbol{y}}
\newcommand{\bw}{\boldsymbol{w}}

\newcommand{\bye}{\boldsymbol{\tilde y}\ssu{E_2}}

\newcommand{\bnu}{\boldsymbol{\nu}}
\newcommand{\I}{\mathcal{I}}

\newcommand{\p}{\mathbf{p}}
\newcommand{\X}{\mathbf{X}}
\newcommand{\mills}{\textsc{mills}}

\newcommand{\ssu}[1]{_{{}_{#1}}}
\title{Composite mixture of log-linear models for categorical data}
\author[1]{Emanuele Aliverti}
\author[2]{David B. Dunson}
\affil[1]{\small\emph{Department of Statistical Sciences, University of Padova}}
\affil[2]{\emph{Department of Statistical Sciences, Duke University}}
\date{}

\begin{document}
\maketitle

\begin{abstract}
	Multivariate categorical data are routinely collected in many application areas.
	As the number of cells in the table grows exponentially with the number of variables, many or even most cells will contain zero observations.
	This severe sparsity motivates appropriate statistical methodologies that effectively reduce the number of free parameters, with penalized log-linear models and latent structure analysis being popular options.
	This article proposes a fundamentally new class of methods, which we refer to as Mixture of Log Linear models (\mills).
	Combining latent class analysis and log-linear models, \mills{} defines a novel Bayesian methodology to model complex multivariate categorical with flexibility and interpretability.
	\mills{} is shown to have key advantages over alternative methods for contingency tables in simulations and an application investigating the relation among suicide attempts and empathy.
\end{abstract}
\section{Introduction}
\label{sec:intro}
 From medical studies to social sciences, there is an immense variety of applications in which the analysis of observations on categorical scales is a routine problem \citep{agresti}.
 Such data can be organized as multiway contingency tables, where individuals are cross classified according to their values for the different variables.  The development of methods to analyse categorical data began well back in the 19th century, and has constantly received attention remaining a very active area of research \citep[e.g.][]{fienberg2007}.  Recent technological developments in data collection and storage motivate novel research questions and innovative methodologies to address them. In particular, it is now standard to collect very high dimensional categorical data in a variety of areas.

Log-linear models are particularly popular for categorical data. Logarithms of cell probabilities are represented as linear terms of parameters related to each cell index, and with coefficients that can be interpreted as interactions among the categorical variables \citep{agresti}.
The relationship between multinomial and Poisson log-likelihoods allows one to obtain maximum likelihood (\textsc{ml}) estimates for log-linear models leveraging standard generalized linear model (\textsc{glm}) algorithms (e.g., Fisher-Scoring), with the vectorized table of cell counts used as a response variable.
As the number of variables increases, the number of cells of the contingency table grows exponentially, many cells will be empty and there will be infinite \textsc{ml} estimates \citep{fienberg2007}.
To overcome this issue and obtain unique estimates, it is often assumed that a large set is coefficients is zero, and estimation is performed via penalised likelihood \citep{nardi2012log,tibshirani2015,wainwright2008graphical,ravikumar2010}.
Since the number of cells is exponential in the number of variables $p$, the computation of the joint cells counts --- required to fit the approaches mentioned above --- becomes unfeasible even for moderate values of $p$; for example, $16$ categorical variables with $4$ categories each define a contingency table with a total number of cells larger than $1$-billion. 

Bayesian approaches for inference in log-linear models often restrict consideration to specific nested model subclasses; for example, hierarchical, graphical or decomposable log-linear models \citep{LauritzenGM}.
Conjugate priors on the model coefficients are available \citep{massam2009conjugate}, but exact Bayesian inference is still complicated since the resulting posterior distribution is not particularly useful, lacking closed form expressions for important functionals -- such as credible intervals -- and sampling algorithms to perform inference via Monte Carlo integration.
As an alternative, the posterior distribution can be analytically approximated with a Gaussian distribution if the number of cells is not excessive \citep{johndrow2018optimal}.
When the focus is on selecting log-linear models with high posterior evidence, stochastic search algorithms evaluating the exact or approximate marginal likelihood are available \citep{dobra2010mode}.
Unfortunately, the size of the model space is enormous and these algorithms scale poorly with the number of variables, being essentially unfeasible in applications with more than $15$ binary variables \citep{johndrow2018optimal}.

A different perspective on analyzing multivariate categorical data relies on latent structures \citep{lazarsfeld1950logical}.
This family of models is specified in terms of one or more latent features, with observed variables modelled as conditionally independent given the latent features.
Marginalising over the latent structures, complex dependence patterns across the categorical variables are induced \citep[e.g.][]{andersen1982latent}.
Representative examples include latent class analysis \citep{lazarsfeld1950logical} and the normal ogive model \citep{lawley1943}, where a univariate latent variable with discrete or continuous support, respectively, captures the dependence structure among the observed categorical variables;  see also \citet[Chapters 9 and 11]{fruhwirth2019handbook} and references therein.
More flexible multivariate latent structures have also been introduced; for example, grade of membership models \citep{erosheva2005comparing} and the more general class of mixed membership models \citep{airoldi2014handbook}.
Specific latent variable models can also be interpreted as tensor decompositions of the contingency tables \citep{Dunson2009,Bhattacharya2012}; see also \citet{kolda2009tensor} for a discussion.

To conduct meaningful and interpretable inferences, it is important for marginal or conditional distributions and measures of association to have a low-dimensional structure.
For example, it is often of substantial interest to characterise bivariate distributions and test for marginal or conditional independence \citep{agresti}.
Leveraging data-augmentation schemes, estimation of latent variable models is feasible in high-dimensional applications \citep[e.g.][]{Dunson2009}; however, these approaches might require many components to adequately characterize complex data, and can lack simple interpretability of the model parameters and the induced dependence structure.
On the other hand, log-linear model directly parameterize  the interactions among the categorical variables \citep{agresti} or the lower-dimensional marginal distributions \citep{bergsma2002marginal}, but estimation is generally unfeasible when the number of variables is moderate to high, due to the huge computational bottlenecks and the massively large model space.
Sparse log-linear models and latent class structures are deeply related in the way in which sparsity is induced in the resulting contingency table \citep{Johndrow2017}, but a formal methodology mixing the benefits of the two model families is still lacking.

Motivated by the above considerations, in this article we introduce a novel class of Bayesian models for categorical data, which we refer to as \textsc{mills}.  We propose to model the multivariate categorical data as a composite mixture of log-linear models with first order interactions, characterising the bivariate distributions with simple and robust models while accounting for dependencies beyond first order via mixing different local models. Such a specification models categorical data with a simple, yet flexible, specification which can take into account complex dependencies with a relatively small number of components. The idea of mixing simple low-dimensional models to reduce the number of parameters needed to characterize complex data has a long history. 
One example is mixing first order Markov models to account for higher order structure \citep{raftery1985model}. See also \citet{fruhwirth2019handbook} for related ideas.

\section{Methods}
\subsection{Log-linear models}
We adopt the notation of \citet{LauritzenGM}.
Let $V=\{1,\dots,p\}$ index a set of $p$ categorical variables. 
Let $(Y_j, j \in V)$ denote variables taking values in the finite set $\I_j$ with dimension $|\I_j|= d_j$. Without loss of generality, we can assume $\I_j = \{1,\dots,d_j\}$.
Categorical data are often collected as an $n \times p$ data matrix with elements $y_{ij} \in \I_j$, $i=1,\dots,n$, $j=1,\dots,p$, and can also be represented as a contingency table.
Let $\I_V = \bigtimes_{j\in V} \I_j$ denote the set with generic element $\bi = (i_1,\dots,i_p)$. 
The elements $\bi$ of $\I_V$ are referred to as the \emph{cells} of the contingency table $\I_V$, which has size $|\I_V| = \prod_{j=1}^p d_j$.
Given a sample of size $n$, the number of observations falling in the generic cell $\bi$ is denoted as $y({\bi})$, with $\sum_{\bi \in \I_V}y({\bi})=n$.

A log-linear model is a generalised linear model for the resulting multinomial likelihood, which represents the logarithms of cell probabilities additively.
Let $\p=(p(\bi), \bi \in \I_v)$ denote the vectorised cell probabilities and let $\btheta$ denote the set of log-linear coefficients.
Following \cite{letac2012bayes,johndrow2018optimal}, it is possible to relate cell probabilities and log-linear coefficients as follows:
\begin{equation}
	\label{logl:matrix}
	\log \p = \X \btheta,
\end{equation}
where $\X$ is a full rank $|\I_V| \times |\I_V|$ matrix if the transformation is invertible; for example, when $\X$ is the identity matrix, the so-called identity parametrisation is obtained.
Identifiability is imposed through careful specification of the matrix $\X$, which determines the model parametrisation and, consequently, constraints on the parameters \citep{agresti}. 
\cref{logl:matrix} can be extended to embrace a larger class of invertible and non-invertible log-linear parametrisations; for example, marginal parametrisations \citep[e.g.][]{bergsma2002marginal, roverato2013log,lupparelli2009}. 

In general, it is desirable to specify a sparse set of $k$ coefficients with $k \ll |\I_v|$, corresponding to some notion of interactions among the categorical variables; for example, representing conditional or marginal independence \citep{agresti}.
When a sparse parameterisation is employed, it is common to remove in \cref{logl:matrix} the columns of $\X$ associated with zero coefficients, thereby obtaining a more parsimonious design matrix with dimension $|\I_V| \times k$.
In this article we focus on the corner parameterisation, which is particularly popular in the literature for categorical data \citep{agresti, massam2009conjugate, letac2012bayes}, and is generally the default choice in statistical software.
The columns of $\X$ under the corner parameterisation can be formally expressed in terms of Moebius inversion \citep[e.g.][Preposition 2.1]{letac2012bayes}; see also \citet[][Lemma 2.2]{massam2009conjugate}. For simplicity in exposition, we prefer to use matrix notation.

Let $\mathbf{y} = ( y(\bi), i \in \I_v)$ denote the vectorised cell counts. The likelihood function associated with the multinomial sampling and log-linear parameters can be expressed, in matrix form, as follows: 

\begin{equation}
	\label{natEF}
	\prod_{\bi \in \I_V} p{(\bi)}^{{y{(\bi)}}} =
	\exp\left\{\mathbf{y}^\intercal \X \btheta - n\kappa(\btheta)\right\} = \exp\left\{\mathbf{\tilde{y}}^\intercal\btheta - n\kappa(\btheta)\right\},
\end{equation}
with $\kappa(\btheta) = \log\left[\mathbf{1}^\intercal \exp(\X\btheta)\right]$.
Such a parametrisation  yields a very compact data reduction, since the canonical statistics $\mathbf{y}^\intercal\X=\mathbf{\tilde{y}}^\intercal$ correspond to the marginal cell counts relative to the highest interaction term included in the model \citep{massam2009conjugate,agresti}.
 In particular, we will consider hierarchical log-linear models which include all the main effects and all the first-order interactions; under such a specification, the canonical statistics $\mathbf{\tilde{y}}$  correspond to the marginal bivariate and univariate tables \citep[e.g][]{agresti}.

 \subsection{Composite likelihood}
 \label{sec:comp}
The log-partition function in \cref{natEF} involves a sum of $|\I_V|$ terms, the total number of cells.  Due to the immense number of cells, the likelihood cannot be evaluated unless $p$ is very small. 
Approximations of intractable likelihoods have been proposed in the literature, with  Monte Carlo maximum likelihood \citep{snijders2002markov,geyer1992constrained} being one option.
Composite likelihoods provide a computationally tractable alternative to the joint likelihood, relying on a product of marginal or conditional distributions; see \cite{varin2011} for an overview.
Extending the work of \citet{meng2013distributed}, \citet{massam2018local} focused on composite maximum likelihood estimation for log-linear models, with a careful choice of the conditional and marginal distributions based on the conditional dependence graph. 
However, the dependence graph is typically unknown and its estimation can be very demanding and affected by large uncertainty \citep{dobra2010mode}.

We propose to replace the joint likelihood with a simple and robust alternative.
Denote as $\mathcal{P}_2$ the set of subsets of $V$ with cardinality $2$.
For each $E_2 \in \mathcal{P}_2$, let $\by_{E_2}$ denote the vectorised  $E_2$-marginal bivariate table of counts. We define, for each $\by_{E_2}$, a saturated log-linear model with corner parametrisation:
\begin{equation}
	\label{biv}
	\p(\by_{E_2};\, \btheta_{E_2}) = \exp\left\{\by_{E_2}^\intercal\X_2 \btheta_{E_2} - n\kappa_2(\btheta_{E_2})\right\} = \exp\left\{\bye^\intercal \btheta_{E_2} - n\kappa_2(\btheta_{E_2})\right\},
\end{equation}
where $\kappa_2(\btheta_{E_2}) = \log\left[\mathbf{1}^\intercal \exp(\X_2\btheta_{E_2})\right]$,  $\dim{\btheta_{E_2}} = \dim{\bye} = |\I_{E_2}| = \prod_{j\in {E_2}} d_j$ and ${\btheta_{E_2}}\in \mathbb{R}^{|\I_{E_2}| }$. 
There is an important difference between $\by_{E_2}$ and $\bye$. The former refers to the $E_2$-marginal bivariate table, while the latter refers to the sufficient statistics of the log-linear model with corner parametrisation, which are  elements of the bivariate and univariate $E_2$-marginal table; see, for example, \citet{agresti}. 

We define a surrogate likelihood function combining the distributions defined in \cref{biv} as 
\begin{eqnarray}
\lefteqn{
	\prod_{E_2 \in \mathcal{P}_2}\p(\by_{{}_{E_2}} ; \btheta_{E_2})^{w_{{}_{E_2}}} } \nonumber \\
&&	=  \exp\left\{\sum_{E_2 \in \mathcal{P}_2}w_{{}_{E_2}}\log \p(\by_{{}_{E_2}} ; \btheta_{E_2})\right\}
=	\exp\left\{\sum_{E_2 \in \mathcal{P}_2} w_{{}_{E_2}} \left[ \bye^\intercal \btheta_{E_2} -n \kappa_2(\btheta_{E_2}) \right] \right\}.
\label{combo}
\end{eqnarray}
Equation (\ref{combo}) is constructed with the same motivation of composing simplified likelihoods from marginal densities in composite likelihood estimation; see, for example, \citet{cox2004note,varin2011}. 
Differently from \citet{massam2018local}, we include contributions for all the bivariate distributions in Equation (\ref{combo}), since the underlying graphical structure is not known a priori, and it is not possible to decide which marginal densities should be included accordingly. Instead, we include all bivariate terms and assign to each component a non-negative weight $w_{{}_{E_2}} \in \mathbb{R}^{+}$, controlling the contribution of the $E_2$ component to the joint likelihood function.

Although it is common to choose unity weights $w\ssu{E_2}=1$ for each ${E_2 \in \mathcal{P}_2}$ \citep[e.g.][]{cox2004note}, careful choice of composite weights can improve efficiency \citep{varin2011}.
Popular choices focus on selecting weights according to some optimality criteria; for example, to correct the magnitude \citep{pauli2011} or curvature \citep{Davison,pauli2011} of the likelihood-ratio test or, more generally, to improve statistical efficiency of the resulting estimating equation \citep[e.g.][]{lindsay2011issues,fraser2019combining,pace2019efficient}.
Beside asymptotic arguments, such procedures are also practically well justified since Equation (\ref{combo}) might include redundant terms, accounting for the same contribution (e.g., marginal univariate) multiple times.
This has motivated the development of more efficient likelihood composition, with the focus on producing sparse estimating equations with few informative components by setting some weights to zero via constrained optimisation \citep{ferrari2016parsimonious,huang2017fast}.
In this article, we build on a similar strategy and aggregate the different components under a Bayesian approach, imposing a sparsity-inducing prior on the weights which favours deletion of redundant terms.

Equation (\ref{combo}) can also be motivated from an inferential point of view.  When interest focuses on inferences for low-dimensional marginal distributions, such as univariates and bivariates, estimates based on the pseudo likelihood in Equation (\ref{combo}) and the original likelihood in \cref{natEF} are equivalent, since the joint model is a closed exponential family which includes only first order interactions in the sufficient statistics \citep[][Theorem 2]{mardia2009maximum}.
With respect to this consideration, it is also worth highlighting that the sufficient statistics $\bye$ of the simplified model in \cref{biv} are actually a subset of the sufficient statistics of the joint model for $\mathbf{\tilde{y}}$ in \cref{natEF} and that $\bigcup_{E_2 \in \mathcal{P}_2} \bye = \mathbf{\tilde{y}}$.

Although in a variety of applications the focus of statistical inference is on low-dimensional margins and related measures of association, Equation (\ref{combo}) may be oversimplified and hence lead to a poor characterisation of multivariate categorical data. For example, there may be significant dependence in the data beyond first order.
To improve flexibility, we propose to use 
Equation (\ref{combo}) to characterize variability within subpopulations using a mixture modeling approach.   
 To formalize this, denote with $\bi_{E_2}$ the elements of $\I_{E_2}$, cells of the $E_2$-marginal bivariate table. 
The contribution for a single observation $y_i=(y_{i1},\dots,y_{ip})$ in Equation (\ref{combo}) can be expressed as 
\begin{equation}
	\label{li}
	\begin{split}
		\tilde \p(y_i ; \btheta, \bw) =	\exp\left\{\sum_{E_2 \in \mathcal{P}_2} w_{{}_{E_2}} \Big[\mathds{1}\left(y_{i}, \bi\ssu{E_2}\right)\X_2 \btheta\ssu{E_2}  -  \kappa_2(\btheta\ssu{E_2}) \Big] \right\},
\end{split}
\end{equation}
with  $\btheta = \{\btheta\ssu{E_2}\}\ssu{E_2\in\mathcal{P}_2}$, $\bw = \{w\ssu{E_2}\}\ssu{E_2\in\mathcal{P}_2}$ and $\mathds{1}(y_{i}, \bi\ssu{E_2})$ 
corresponding to a vector of length $|\I_{E_2}|$ with a $1$ in the position for the cell in which the $E_2$ component of $y_{i}$ falls and all other elements $0$.
We introduce a latent group indicator $z_i \in\{1,\dots,H\}$ with $\mbox{pr}[z_i = h] = \nu_h$, indexing the subpopulation for the $i$th subject.
We use Equation (\ref{combo}) as a local model for characterizing the dependence structure of subjects in the same latent group. 
By allowing the weights $w\ssu{E_2}$ to vary across subpopulations, we allow the complexity of the local model to vary substantially and adapt to the subpopulation-specific structure.

Considering only observations belonging to group $h$ and denoting with ${n_h = \sum_{i=1}^n \mathds{1}[z_i = h]}$ the number of units in group $h$, we interpret Equation (\ref{combo}) as a model for the contingency table conditional on group membership, as
\begin{equation}
	\label{comboH}
	\tilde \p(\by^h ; \btheta^h, \bw^h \mid \mathbf{z} = h) \	=	\exp\left\{\sum_{E_2 \in \mathcal{P}_2} w\ssu{E_2}^h\Big[ \bye^{h\intercal} \btheta\ssu{E_2}^h -n_h  \kappa_2(\btheta\ssu{E_2}^h)\Big]\right\},
\end{equation}
where the composite likelihood weights $\bw^h = \{w\ssu{E_2}^h\}_{E_2 \in \mathcal{P}_2}$ and the log-linear parameters $\btheta^h = \{\btheta_{E_2}^h\}_{E_2\in\mathcal{P}_2}$ are allowed to vary across mixture components to characterise different dependence patterns in different subpopulations.
Marginalising over the latent feature $\mathbf{z}$ and considering the contribution for all the data points, we obtain a joint model with likelihood function equal to
\begin{equation}
	\label{model}
	\tilde \p(\by ; \btheta,\bw,\bnu) = \prod_{i=1}^n \sum_{h=1}^H\nu_h\, \tilde \p(y_i ; \btheta^h, \bw^h),
\end{equation}
with $\btheta = \{\btheta^h\}_{h=1}^H$, $\bw = \{\bw^h\}_{h=1}^H$ and $\bnu=\{\nu_h\}_{h=1}^H$. 

	The adaptive log-linear structure imposed within each component of \Cref{comboH} allows one to characterize complex dependence patterns with few components.
Increasing the number of components $H$, any structure can be effectively characterised under \textsc{mills}.
The following Lemma formalizes the ability of \textsc{mills} to represent any 
$\p \in \mathcal{S}\ssu{|\mathcal{I}_V|}$, with $\mathcal{S}\ssu{|\mathcal{I}_V|}$ denoting the $(|\mathcal{I}_V| - 1)$-dimensional simplex. See the Appendix for a proof.

\begin{lemma}
	\label{lemma1}
	Any $\p \in \mathcal{S}\ssu{|\mathcal{I}_V|}$ admits representation (\ref{model}) for some $H$, with $\nu_h \in (0,1)$ such that $\sum_{h=1}^H \nu_h = 1$.

\end{lemma}

Equation (\ref{model}) provides a compact model for efficiently making inference on low-dimensional marginals. For example, a natural estimate for the $E_2$ bivariate distribution is given by
\begin{equation*}
	\hat{\mbox{pr}(\bi_{E_2})} = \sum_{h=1}^H\nu_h\,\exp\left\{\X_2 \btheta_{E_2} - \kappa_2(\btheta_{E_2})\right\},
\end{equation*}
which corresponds to a weighted average of local estimates, with weights given by the mixture weights.

\subsection{Prior specification}
We proceed with a Bayesian approach to inference, and specify prior distributions for the parameters $\bnu$, $\btheta_{E_2}^h$ and $\mathbf{w}$. We rely on Dirichlet and Gaussian distributions, letting
\begin{equation}
	\label{prior}
	(\bnu \mid H) \sim \mbox{\textsc{dir}}\left(\frac{1}{H}, \dots, \frac{1}{H}\right), \quad ( \btheta_{E_2}^h \mid \sigma^2) \stackrel{\text{iid}}{\sim} \mbox{N}_{|\I_{E_2}|}(\mu\ssu{E_2},\sigma^2I),\quad E_2 \in \mathcal{P}_2,\quad h=1,\dots,H.
\end{equation}
Estimation for the number of active components is performed by choosing a conservative upper bound $H_0$ for $H$, and specifying a sparse Dirichlet distribution on the mixture weights to automatically favour deletion of redundant components \citep{rousseau2011asymptotic}.
The Gaussian priors on the log-linear parameters allow simple inclusion of prior information, for example reflecting knowledge on the expected direction and strength of the association between pairs of variables.
Moreover, computations are particularly easy adapting the P\`{o}lya-Gamma data-augmentation strategy for the multinomial likelihood and Gaussian prior \citep{polson2013bayesian}.
Under an exponential family representation, other conjugate priors are available for the natural parameters \citep[e.g.][]{massam2009conjugate, bradley2019bayesian}.
However, Gaussian priors have simpler interpretation and facilitate computation.

As motivated in \cref{sec:comp}, the prior distribution for the composite weights $w\ssu{E_2}^h \in \mathbb{R}^{+}$ should induce sparse configurations, deleting redundant components.
To address this with computational tractability, we rely on a continuous spike and slab prior.
Such a strategy focuses on introducing latent binary indicators $\delta\ssu{E_2}^h \in \{0,1\}$ encoding exclusion or inclusion of the $E_2$ component in \Cref{combo}, with $\mbox{pr}[\delta\ssu{E_2}^h = 1 \mid \gamma_0^h] = \gamma_0^h$. 
Conditionally on $\delta\ssu{E_2}^h$, each $w\ssu{E_2}^h$ is drawn independently either from a distribution concentrated around zero, $P_0$, or from a diffuse distribution over the real positive line, which we denote as $P_1$.
For computational convenience, we rely on the following hierarchical specification for $w\ssu{E_2}^h$.
\begin{equation}
	\begin{split}
	\label{eq:ssprior}
	(\delta\ssu{E_2}^h \mid \gamma_0^h) &\stackrel{\text{iid}}{\sim} \mbox{\textsc{bernoulli}}(\gamma_0^h) \\
	(w\ssu{E_2}^h \mid \delta\ssu{E_2}^h) &\stackrel{\text{iid}}{\sim} \mbox{\textsc{gamma}}(1+a_0^h\delta\ssu{E_2}^h, a_1^h), \quad E_2 \in \mathcal{P}_2,\, h = 1,\dots, H
\end{split}
\end{equation}

Although it is possible to replace the spike with a Dirac mass at $0$, we follow \citet{ishwaran2005spike}, and introduce a continuous shrinkage prior, which is shown to generally improve computation and mixing; see also \citet{legramanti2019bayesian} for related arguments.

\begin{figure}
       \centering
       \includegraphics[width = \textwidth]{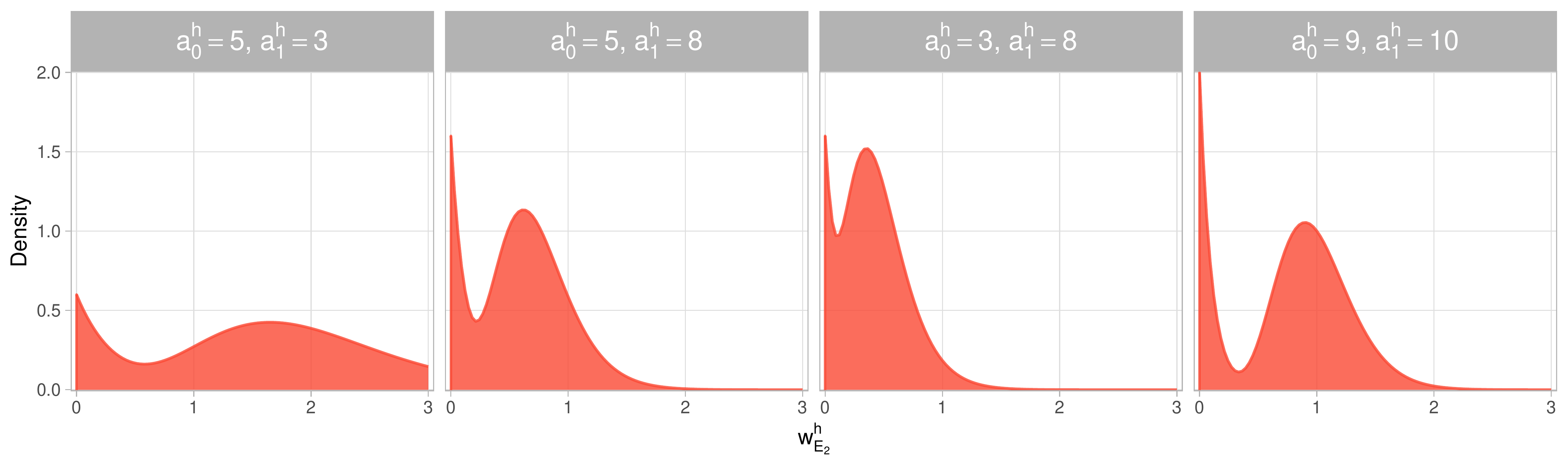}
       \caption{Graphical illustration of the prior distribution of \Cref{eq:ssprior} for different  hyper-parameter values. In each panel, $\gamma_0^h=0.2$.}
       \label{fig:prior}
\end{figure}

Marginalising out $\delta\ssu{E_2}^h$ from \Cref{eq:ssprior}, we obtain a discrete mixture between a Gamma distribution with shape $1$ and rate $a_1^h$ (Exponential), and a Gamma distribution with shape $(1+a_0)$ and rate $a_1^h$.
The parameter $\gamma_0$ controls the prior proportion of active terms, and is assigned  a symmetric $\mbox{\textsc{beta}}(0.5,0.5)$ prior \citep{ishwaran2005spike}.
Specifying large values for $a_1^h$, substantial mass around $0$ is induced, while $a_0^h$ controls the mean and variance for the Gamma distribution associated with the slab.
See \Cref{fig:prior} for a graphical illustration of the prior density over illustrative combinations of hyper-parameters.
In the absence of explicit prior information on the composite likelihood weights, we recommend to elicit the prior distribution to include values around $1$ with high probability in the slab component.
Such choice guarantees that, when a component is included, default units weights are selected with high probability a priori, centering the model around a standard specification.

\subsection{Posterior computation}

There is a rich literature on the use of alternative likelihoods for Bayesian inference; for example, approximate likelihood \citep{Efron1993},  partial likelihood \citep{Raftery95accountingfor}, empirical likelihood \citep{Lazar2003} and adjusted profile likelihood \citep{Chang2006}, among many others. See also \citet{Greco2008} for related arguments.
Although the use of composite likelihoods in Bayesian inference is more recent \citep[e.g.][]{Davison,pauli2011}, it has received substantial attention \citep{miller2019}.
Related to these approaches, we conduct inference using the composite posterior distribution %
\begin{equation}
	\label{eq:post}
	\tilde \pi(\btheta,\bnu \mid \by) \propto \pi(\btheta)\pi(\bnu)\pi(\bw)\tilde \p(\by ; \btheta,\bw,\bnu).
\end{equation}
It is important to show that this composite posterior is a proper probability distribution, which is not guaranteed for general composite-type likelihoods.  However, the following Lemma shows that our composite posterior does have this property. See the Appendix for a proof.
\begin{lemma}
	\label{lemma:prob}
$\tilde \pi(\btheta,\bnu \mid \by)$ is a proper probability distribution.
\end{lemma}
To make inference from \Cref{eq:post}, we rely on an \textsc{mcmc} algorithm whose main steps are described in \Cref{alg:gibbs}.
We leverage the P\`{o}lya-Gamma data augmentation strategy of \citet{polson2013bayesian} to obtain conditionally conjugacy between the Gaussian prior and the multinomial likelihood, while the mixture weights $\bnu$ and composite weights  $\bw$ are updated sampling from Dirichlet and Gamma full conditional distributions, respectively.
Similarly, the mixture indicator $z_i$ is sampled from its full conditional categorical distribution, for each $i=1, \dots,n$.
The main bottleneck is storage of the conditional bivariate terms, which have size $\mathcal{O}(Hp^2d^2)$.
Although the introduction of the spike and slab strategy drastically improves estimation --- since many components are effectively assigned to zero weight at each iteration and Equation (\ref{combo}) involves only few informative components --- the storage of redundant terms is required during estimation and can be burdensome in large $p$ applications.  The proposed algorithm easily scales up to a motivating application involving $p=64$ items with $d=5$ categories each, relying on a mixed \textsc{r} and \textsc{c++} implementation on a standard laptop; see \Cref{sec:app}.
Scaling to much larger cases can potentially be accomplished by replacing the continuous spike with a mass at zero or thresholding redundant components as an approximation.  

\RestyleAlgo{boxruled}
\begin{algorithm}[tb]
	\caption{One cycle of Gibbs sampler for \textsc{mills}.}
\label{alg:gibbs}
\For{$h = 1, \dots, H$} {
	\For{$E_2 = 1, \dots, |\mathcal{P}_2|$} {
It is convenient to reparametrize the \textsc{mills} likelihood as $\widetilde{\btheta}\ssu{E_2}^h = \X_2\btheta\ssu{E_2}^h$, corresponding to the cell-specific multinomial log-odds.
		The Gaussian prior on $\btheta\ssu{E_2}^h$ induces a Gaussian prior on $\widetilde{\btheta}\ssu{E_2}^h$ with covariance matrix $\X_2^\intercal\X_2$. Therefore, the prior precision of each element of $\widetilde{\btheta}\ssu{E_2}^h$ given the others is given by the diagonal elements of $(\X_2^\intercal\X_2)^{-1}$.

		Sample each $\widetilde{\btheta}\ssu{E_2}^h$ from a conditionally-conjugate Gaussian distribution, adapting the P\`{o}lya-Gamma strategy to the multinomial likelihood \citep{polson2013bayesian}.
}
}
\For{$h = 1, \dots, H$} {
	\For{$E_2 = 1, \dots, |\mathcal{P}_2|$} {
Sample each $\delta\ssu{E_2}^h$ from a Bernoulli distribution with probability of success equal to 
		\begin{equation*}
			\frac{\gamma_0^h\mbox{\textsc{gamma}}(w^h\ssu{E_2}; 1 + a_0^h, a_1^h - \ell^h\ssu{E_2}) }{\gamma_0^h\mbox{{\textsc{gamma}}}(w^h\ssu{E_2}; 1 + a_0^h, a_1^h - \ell^h\ssu{E_2}) + ( 1- \gamma_0^h)\mbox{{\textsc{gamma}}}(w^h\ssu{E_2}; 1, a_1^h - \ell^h\ssu{E_2}) },
			\end{equation*}
with $\ell\ssu{E_2}^h =\log[ \bye^{h\intercal} \btheta\ssu{E_2}^h -n_h  \kappa_2(\btheta\ssu{E_2}^h)\Big]$ and with $\mbox{{\textsc{gamma}}}(x;a,b)$ denoting the density of a Gamma distribution with shape $a$, rate $b$ evaluated in $x$.
Note that $\ell\ssu{E_2}^h$ is always negative, and therefore there is no ambiguity in the evaluation of the Gamma density.
}

	\For{$E_2 = 1, \dots, |\mathcal{P}_2|$} {
		Sample the composite weight $w\ssu{E_2}^h$ from 
		\begin{equation*}
		\mbox{\textsc{gamma}}\left(1+a_0^h\delta\ssu{E_2}^h, a_1^h - \ell^h\ssu{E_2}\right)
	\end{equation*}
}
Sample the slab probability $\gamma_0^h$ from a
\begin{equation*}
	\mbox{\textsc{beta}}\left(\dfrac{1}{2} + \sum\ssu{E_2 \in \mathcal{P}_2} \delta\ssu{E_2}^h, \dfrac{1}{2} + |\mathcal{P}_2| - \sum\ssu{E_2 \in \mathcal{P}_2} \delta\ssu{E_2}^h\right)
\end{equation*}}

\For{$i = 1, \dots, n$} {
	Sample $z_i$ from
\begin{equation*}
	\mbox{\textsc{categorical}}\left(\dfrac{\nu_1 \tilde \p(y_i ; \btheta^1, \bw^1)}{\sum_{h=1}^H \nu_h \tilde \p(y_i ; \btheta^h, \bw^h)}, \cdots, \dfrac{\nu_H \tilde \p(y_i ; \btheta^H, \bw^H)}{\sum_{h=1}^H \nu_h \tilde \p(y_i ; \btheta^h, \bw^h)}  \right)
\end{equation*}
	with $\p(y_i ; \btheta^h, \bw^h)$ defined in \Cref{li}.
}
Sample $\bnu$ from
\begin{equation*}
	\mbox{\textsc{dirichlet}}\left(n_1 + \dfrac{1}{H}, \cdots, n_H + \dfrac{1}{H}\right),
\end{equation*}
with $n_h =  \sum_{i=1}^n \mathds{1}[z_i = h]$.
\end{algorithm}

\section{Simulation Study}

In order to evaluate the model performance, we considered a simulation study over four different settings.
In each scenario, we focus on an artificial sample of size $n=400$, with $p=15$ categorical variables and $d_1 = \dots \ d_{15}=4$ categories.
In the first scenario, multivariate categorical data are generated from a latent class model with $H = 5$ components and probabilities generated from a uniform prior on the simplex. 
The second scenario samples categorical variables $j \in \mathcal{J} = (1,2,3,4,5)$ from a dense log-linear model with first order interactions and coefficients randomly sampled from a Gaussian distribution with standard deviation $0.1$, while the remaining categorical variables $j \notin \mathcal{J}$ are generated from independent Dirichlet-Multinomial distributions with hyper-parameter $(3,3,3,3)$. 
In the third scenario, we focus on the same groups of variables, imposing more structure on the variables in the group $\mathcal{J}$, which are sampled from the joint probability mass function assigning probability $0.1$ to the cells $\bi_{\mathcal{J}} \in \{ (1, \dots, 1),\dots, (4, \dots, 4)\}$ and probability $0.6$ to the remaining cells in equal proportion; see also \citet{russo2018}.
The remaining variables $j \notin \mathcal{J}$ are generated from independent Dirichlet-Multinomial distributions with hyper-parameter $(3,3,3,3)$.
The fourth and last scenario further complicates the second one by introducing an additional group of variables $\mathcal{J}^\prime = (5,6,7,8,9,10)$, generated from a dense hierarchical log-linear model with first and second order interactions, and coefficients randomly sampled from a Gaussian distribution with standard deviation $0.1$. 

\begin{figure}[bt]
	\centering
	\includegraphics[width = \textwidth]{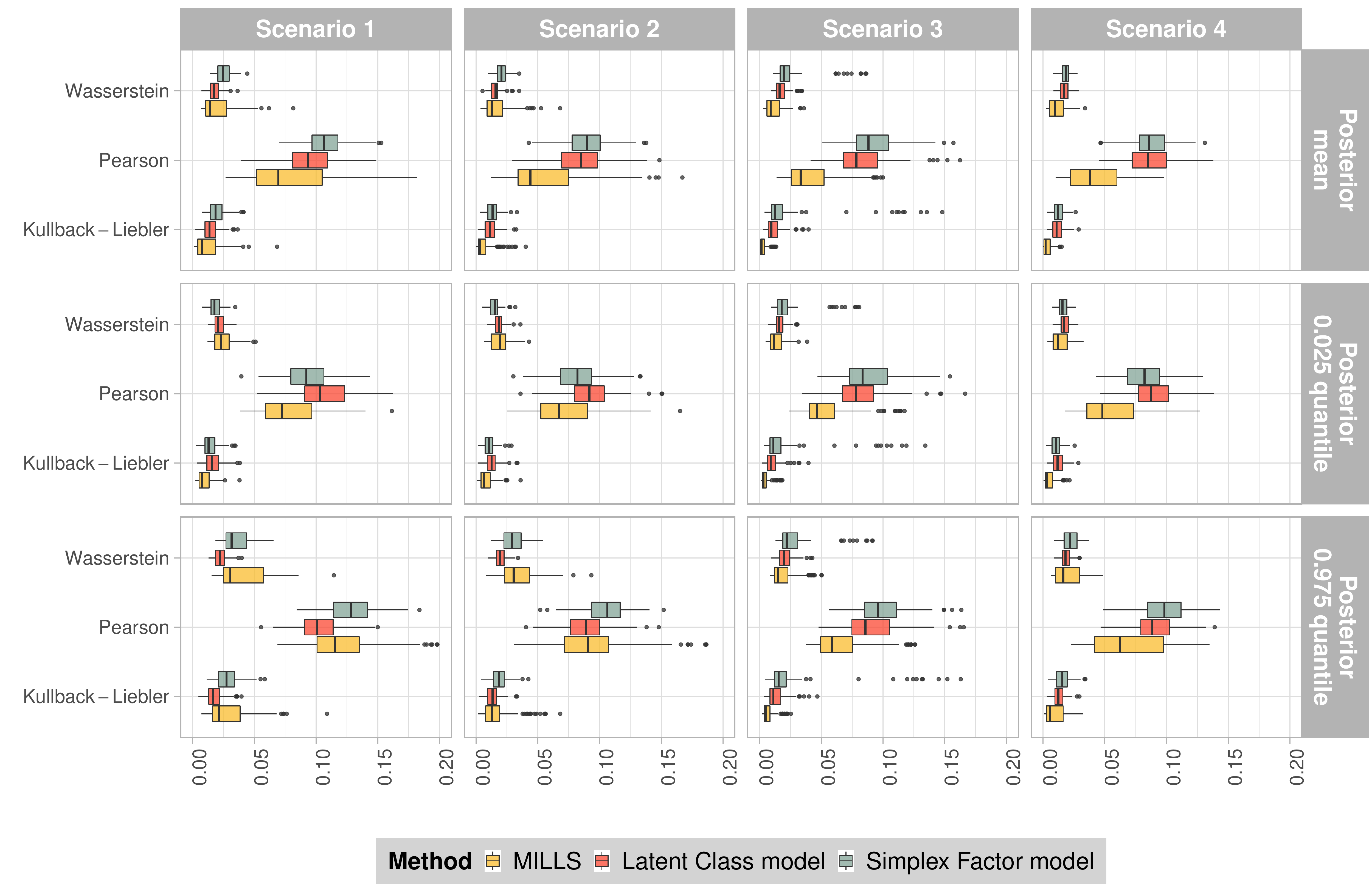}
	\caption{{Simulation studies. Wasserstein distance, normalised Pearson's residuals and absolute Kullback-Leibler divergence between estimates and observed quantities.
		First row refers to posterior means; second and third to posterior $0.025$ and $0.975$ quantiles, respectively.
Yellow boxplots refer to \textsc{mills}. Red and gray to Latent Class model and Simplex Factor model, respectively.}}
	\label{fig1}
\end{figure}
The focus of these settings is on inducing challenging data generating processes, characterised by heterogeneous dependence across subsets of categorical variables.
Posterior inference for \textsc{mills} relies on $1000$ iterations collected after a burn-in period of $1000$, setting a conservative upper bound $H=5$ and $\sigma_2=3,a_0=10,a_1=10$. 
Trace plots and \textsc{mcmc} diagnostics indicate good mixing in all the settings considered.
As competitor approaches, we considered two flexible latent variable models, whose estimation is feasible in the settings under investigation.
The first is a Bayesian specification of a latent class model with $H=10$ classes, sparse Dirichlet priors over the mixture weights and unit Dirichlet priors on the class-specific probabilities. 
Such an approach corresponds to a finite mixture of product multinomial distributions; see, for example, \citet[Chapter 9]{fruhwirth2019handbook} for an introduction.
The second competitor is a simplex factor model \citep{Bhattacharya2012} with $H=10$ latent factors, which provides a mixed membership model \citep[e.g.][]{airoldi2014handbook} for multivariate categorical data.
Again, we rely on a Bayesian specification relying on independent Dirichlet priors over the model parameters.
As outlined in \Cref{sec:intro}, both approaches induce a parsimonious low-rank decomposition of the probability mass function, and the connection between such decompositions and a log-linear model specification has been explored in \citet{Johndrow2017}.

The focus of the simulations is on evaluating the ability of the approaches in estimating low-dimensional functionals of the data. We focus on the set $\mathcal{P}_2$ of bivariate distributions, whose precise estimation is crucial for computing measures of bivariate associations and making inference on the dependence structure. 
\Cref{fig1} illustrates the variability across $\mathcal{P}_2$ under the four simulations settings and for the three approaches considered.
The first row of \Cref{fig1} shows estimated posterior mean for the three methods, compared with their empirical counterparts in terms of Kullback-Leibler divergence, Wasserstein distance and normalised Pearson's residuals.

The first column of \Cref{fig1} illustrates results for the first scenario, and suggests that when data are generated from a latent class model, the three approaches are comparable in terms of goodness of fit, with \textsc{mills} resulting in predictions which are more accurate on average, but also more variable.
The good performance of the latent class model was expected, since such an approach is correctly specified in the first scenario.
As outlined in \Cref{sec:comp}, \textsc{mills} can induce a latent class specification as a special case, and therefore its performance is on average similar with the competitors, but also characterized by a higher variability which might be due to the estimation of the richer dependence structure imposed within each mixture component.
In the second and third scenario, results indicate the superiority of \textsc{mills} with respect to the latent class model and the simplex factor model.
Such a result highlights the ability of the proposed approach to adapt to settings with heterogeneous dependence patterns across subsets of variables; the third column of \Cref{fig1}, in addition, confirms how \textsc{mills} achieves better performance than the competitors also when such dependence patterns go beyond first order interactions.
Lastly, the fourth scenario illustrates the ability of \textsc{mills} to adapt better than the competitors to highly complex settings, dependence patterns beyond first order interactions and involving multiple sub-groups of variables.
The superiority of \textsc{mills} in such settings might be due to the parsimonious composite likelihood specification of Equation (\ref{combo}), with adaptive estimation of the degree of dependence required by each component.
Variability in the simulations is assessed considering the posterior $0.025$ and $0.975$ quantiles of the estimated bivariate distributions, graphically reported for each method in the second and third row of \Cref{fig1} respectively.
	The main empirical findings are consistent with the discussion outlined above, indicating an overall better performance of \textsc{mills} under complex data generating processes.

\section{Application}
\label{sec:app}
We consider a psychiatric study of suicide attempts \citep[e.g.][]{nock2008,deleo2004}.
Studies on survival of suicide attempts are crucial for the development of novel intervention treatments based on the early identification of psychological symptoms \citep[e.g.][]{hawton1988}, and also for accurate descriptions of the psycho--pathological profiles of individuals more likely to conduct suicidal acts.
For example, depression and hostility symptoms are often associated with suicide attempts \citep{ben2004}, while some recent work has  suggested that empathy could be an important risk factor associated with specific psychiatric disease and the suicidal act \citep[e.g.][]{lachal2016}.

It is of interest to analyse the psychopathology of suicide attempt patients, their empathic profile and the possible interactions across these two psychological aspects.
Individuals analysed in the study correspond to a sample of $58$ inpatients hospitalized after an attempted suicide at the psychiatric ward of Padova Hospital (Italy) between January 2017 and December 2018.  
Data were collected by self administered questionnaires aimed at evaluating different psychological aspects of attempted suicide, with the Symptom Check List (\textsc{scl-90}) \citep{derogatis1973} and the Interpersonal Reactivity Index (\textsc{iri}) \citep{davis1980} being reliable instruments for these purposes.

Specifically, the \textsc{scl-90} is commonly used to describe psychiatric symptoms, using $90$ items scored on a five-point Likert scale; additionally, scores can be grouped into nine subscales (somatization, obsessive-compulsive, interpersonal sensitivity, depression, anxiety, hostility, phobic anxiety, paranoid ideation, psychoticism) corresponding to well-defined psychiatric profiles \citep{derogatis1973}.
As suggested by our clinician collaborators, it is of particular interest to focus on $4$ subscales of the questionnaire: obsessive-compulsive (\textsc{oc}), depression (\textsc{dep}), anxiety (\textsc{anx}) and hostility (\textsc{hos}), encompassing a total of $28$ items. See \Cref{app:sclQ} in the Appendix for an illustration of the items under investigation.

The \textsc{iri} is a $28$-item instrument scored on a five-point Likert scale that measures the emotional and cognitive components of a person's empathy, with four subscales.
The \textsc{iri} measures the cognitive capacity to see things from the point of view of others (Perspective Taking, \textsc{pt}), the tendency to experience reactions of sympathy, concern and compassion for other people undergoing negative experiences (Empathic Concern, \textsc{ec}), the tendency to experience distress and discomfort in witnessing other people's negative experiences (Personal Distress, \textsc{pd}) and the capacity to strongly identify oneself with fictitious characters in movies, books, and plays (Fantasy, \textsc{fs}).
For a detailed illustration of the  items, see \Cref{app:iriQ} in the Appendix.

In the psychological literature, investigation of the relationship among different empathic profiles and psycho-pathological symptoms has been a challenging research objective.
Generally, variations in empathy are also associated with depression \citep{cusi2011,schreiter2013}, obsessive compulsive disorders \citep{fontenelle2009}, anxiety \citep{perrone2014} and  hostility \citep{guttman2002}.
For example, a frequent symptom of depression is the inability to perceive our own emotions, which is also realistically associated with the inability to comprehend other individuals' emotions \citep[e.g.][]{cusi2011}.
Another example includes anxiety symptoms, which are likely to be associated with personal distress and hostility \citep{guttman2002}.
However, the relationship among psycho-pathological symptoms and empathic profiles in patients attempting suicide is still not completely understood.
Indeed, individuals who attempted suicide might exhibit unexpected association patterns across the psycho-pathological diseases and empathic profiles.

\begin{figure}[bt]
	\centering
	\includegraphics[width = \textwidth]{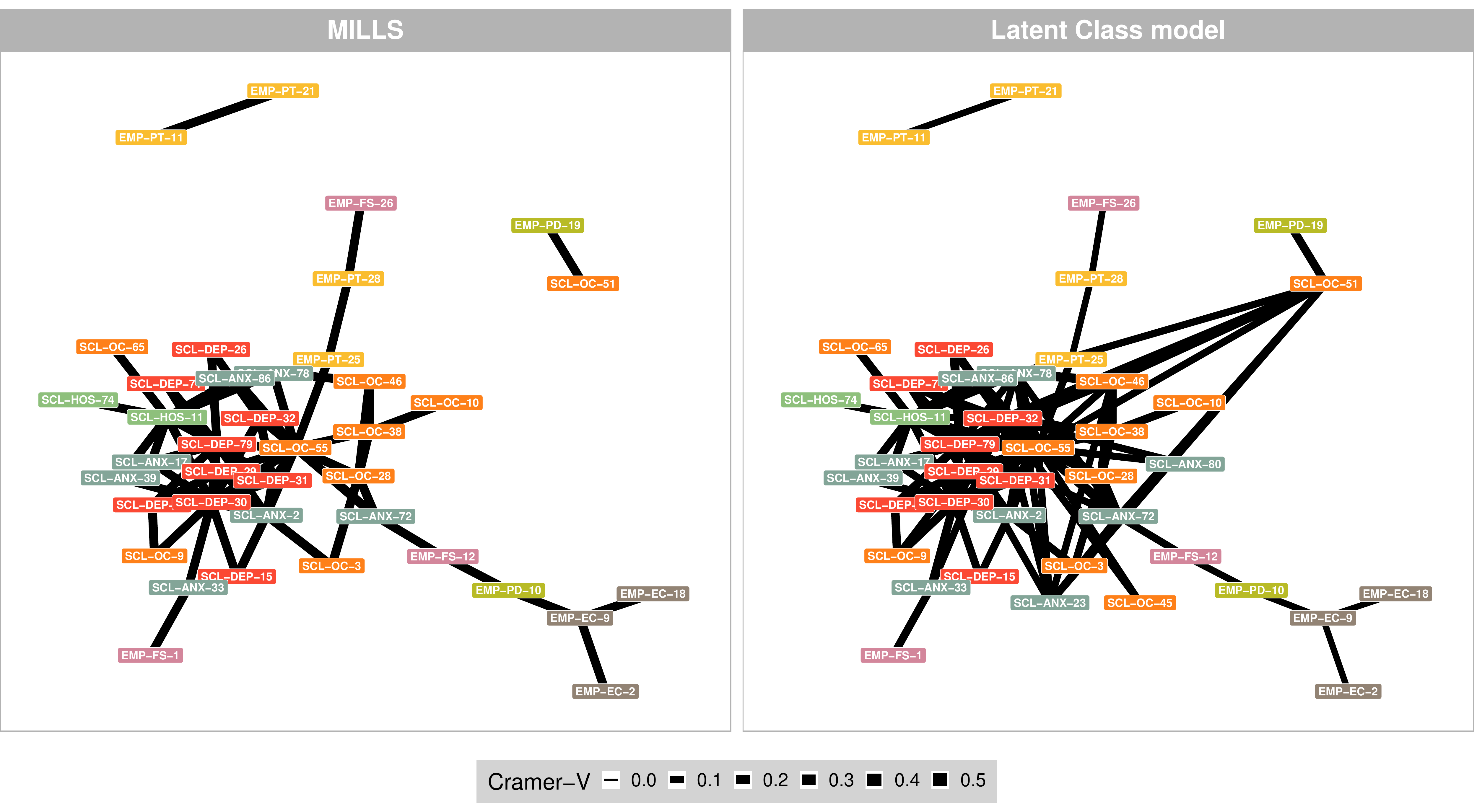}
	\caption{Association structure of the items. Color of the nodes varies with subscales, while edge widths vary with the value of the posterior mean of the pairwise Cramer-V.}
	\label{fig:app}
\end{figure}

Posterior inference for \textsc{mills} uses the same specification as in the simulations, relying on $3000$ iterations collected after a burn-in of $1000$.
Posterior computation is based on an \textsc{r} implementation and requires approximately $7$ minutes per $100$ iterations and $4$\textsc{gb} of \textsc{ram} on a laptop with an \textsc{intel(r) core(tm)} \textsc{i7-7700hq} @ 2.8 \textsc{ghz} processor running Linux.
	We conducted sensitivity analyses for different hyper-parameter specifications, replicating posterior computation with values $a_0 \in \{10,100, 1000\}, a_1 \in \{10,100, 1000\}$ and $\sigma_2 \in \{3,10\}$. The overall empirical findings were robust across changes in hyper parameters. 

Posterior inference focuses on bivariate associations measured via the Cramer-V, which can be easily computed via Monte Carlo integration leveraging the \textsc{mcmc} output.
\Cref{fig:app} illustrates the dependence structure as a graph, with nodes corresponding to the categorical variables and edges to their associations, with thicker edges corresponding to stronger associations and higher Cramer-V. 
The left panel of \Cref{fig:app} refers to \textsc{mills}, and the right panel to a latent class model with $H=10$ components and the same specification as in the simulations.

Our empirical findings highlight the presence of strong associations across several sub-scales, in particular within items associated with similar profiles.
For example, the bulks of red (\textsc{scl-dep}) and orange (\textsc{scl-oc}) nodes in \Cref{fig:app} denote items associated with depressive and obsessive compulsive profiles, respectively, suggesting significant interconnections within these two sub-scales.
To some extent, this result confirms the validity of the tools to measure psycho-pathological symptoms, which characterize consistent psychological profiles and highlights that such  profiles are strongly associated in suicide attempt survivors.
In addition, some items corresponding to different profiles measured within the same questionnaire are characterized by strong interactions.
For example, the empirical findings indicate an association between an anxious subject  \textsc{scl-anx-2} (\emph{``Nervousness or shakiness inside''}) and \textsc{scl-dep-15} (\emph{``Thoughts of ending your life''}) in suicide attempt survivors.

\begin{figure}[bt]
	\centering
	\includegraphics[width = \textwidth]{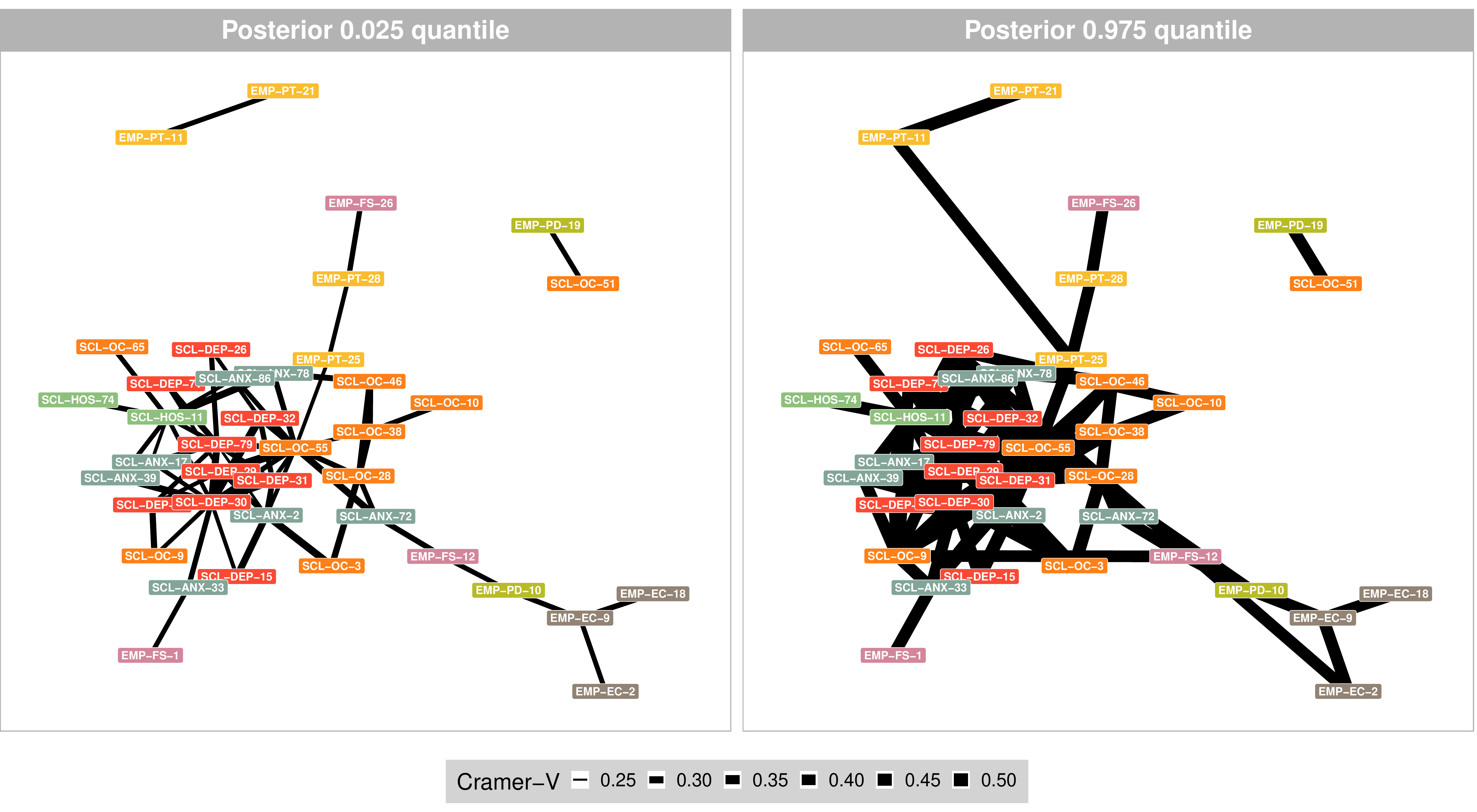}
	\caption{Posterior quantiles of the pairwise Cramer-V under \textsc{mills}}
	\label{fig:lo}
\end{figure}

Other interesting associations involve items in different subscales. 
For example, there is an association between an item from the \textsc{iri} questionnaire \textsc{emp-fs-1} (\emph{``I daydream and fantasize, with some regularity, about things that might happen to me''}) with the item \textsc{scl-anx-33} (\emph{``Feeling fearful''}), and also the \textsc{scl-dep-30} item (\emph{``Feeling blue''}).
This dependence structure is coherent with a paranoid profile, with fantasies about things that might happen and with such thoughts inducing substantial fear and sadness.
Another interesting association involves the items \textsc{scl-ic-51} (\emph{``Your mind going blank''}) and \textsc{iri-19} (\emph{``I am usually not effective in dealing with emergencies.''}), which are consistent with a profile with low-capacity to handle complex situations with calm.
Panels of \Cref{fig:lo} assess uncertainty in \textsc{mills} estimation considering the $0.025$ and $0.975$ posterior quantiles of the Cramer-V, and suggesting that the estimated structure is maintained considering such posterior summaries.

Results from a Latent Class model -- reported in the right panel of \Cref{fig:app} -- are consistent with inference based on \textsc{mills}, suggesting dense associations among items related to the same psychopathologies.
When data exhibit complex dependence and the sample size is small compared with the dimensionality of the problem -- as in this application, with $n=56$ subjects -- the latent class model often has unsatisfactory performance and detects many false signals \citep{Zhou2015}.

\section{Discussion}

This article has proposed a new approach for the analysis of categorical data relying on a mixture of log-linear models, with a computationally convenient composite likelihood-type specification facilitating implementation.  Although multivariate categorical data are very commonly collected in many different areas, we still lack methods for doing inferences on associations among variables in a flexible manner that can accommodate more than a small number of variables.  Current log-linear models do not scale up to large contingency tables and latent structure methods sacrifice some of the key advantages of log-linear models in terms of providing a direct and interpretable model on the association structure.  Hence, latent structure models are in some sense too black box and unstructured, potentially leading to a non-parsimonious characterization of the data, and necessitating a moderately large number of latent components.

The goal of the proposed framework is to borrow the best of both worlds between latent structure and log-linear models.  We have illustrated through a detailed case study to data on suicide attempts that the proposed framework can have a practically relevant impact on inferences for important real world applications. Routine implementation of the proposed approach is facilitated by the availability of an \textsc{r} package, refer to \url{github.com/emanuelealiverti/mills} having default hyper-parameter values for routine use.  There are many interesting next steps in terms of including further computational simplifications to facilitate scaling up, and to include more complex data structure; for example, having missing data, mixed measurement scales, etc.

\section{Acknowledgments}
The case study illustrated in \Cref{sec:app} has been motivated by a collaboration with doctor Paolo Scocco from Padova Hospital, which is kindly acknowledged for providing the data and the stimulating discussions. Emanuele Aliverti would also like to acknowledge Prof. Giovanna Capizzi, Massimiliano Russo and Daniele Durante for the stimulating discussions on the first draft of this work.
A preliminary version of this work was included in the Ph.D. dissertation of Emanuele Aliverti.
This work was partially funded by \textsc{miur-prin} 2017 project 20177BR-JXS, as well as grant R01ES027498 of the National Institute of Environmental Health Sciences of the United States Institutes of Health, and grant N00014-16-1-2147 from the United States Office of Naval Research.
\appendix

\section{Appendix}
\begin{proof}[Proof of \Cref{lemma1}]
	The proof for the full generality of \textsc{mills} relies on illustrating how such a specification induce a finite mixture of independent multinomial distributions as a special case.
	Without loss of generality, consider equal number of categories $d_j=d$ for $j=1, \dots,p$ and equal weights $\bar{w}\ssu{E_2}^h = 1/(p-1)$   for $E_2 \in \mathcal{P}_2$ and $h=1,\dots,H$. 
	Introduce a set of constrained log-linear coefficients $\bar{\btheta}\ssu{E_2}^h$ as $\bar{\btheta}\ssu{E_2}^h = \mathbf{L} \otimes \btheta\ssu{E_2}^h$, where $\mathbf{L}$ denotes a vector of length $d^2$ with the first $1+p(d-1)$ elements equal to $1$ and the remaining $0$, and $\otimes$ denoting element-wise product.
	Therefore, each $\bar{\btheta}\ssu{E_2}^h$ induces a log-linear independence model, which includes only main effects.
	Under the above constraints, 
	\begin{equation}
		\sum_{h=1}^H \nu_h \exp\left\{\sum_{E_2 \in \mathcal{P}_2} \bar{w}\ssu{E_2}^h \Big[\X_2 \bar{\btheta}\ssu{E_2}^h  -  \kappa_2(\bar{\btheta}\ssu{E_2}^h) \Big] \right\},
	\end{equation}
	corresponds to a discrete mixture of product multinomial distribution, for which Theorem 1 of \citet{Dunson2009} follows directly, after noticing that 
	\begin{equation}
		\boldsymbol{\psi}_h^{(j)} = \mathbf{M} \prod_{\substack{E_2 \in \mathcal{P}_2 : j \in E_2}} \left[\exp\Big(\X_2 \bar{\btheta}\ssu{E_2}^h  -  \kappa_2(\bar{\btheta}\ssu{E_2}^h) \Big)\right]^{\bar{w}\ssu{E_2}^h},
\end{equation}
where $\mathbf{M}$ denotes a $d \times d^2$ marginalisation matrix, comprising zeros and ones in appropriate positions \citep[e.g.][]{lupparelli2009}.
\end{proof}
\begin{proof}[Proof of \Cref{lemma:prob}]
	In order to show that is a proper probability distribution, it is necessary to show that the normalising constant is finite, which correspond to showing that
\begin{equation}
	\label{bound}
{\int\int  \pi(\btheta)\pi(\bnu)\pi(\bw)\tilde \p(\by ; \btheta,\bw,\bnu) \mbox{d}\btheta\mbox{d}\bnu\mbox{d}\bw}
	\int\int  \pi(\btheta)\pi(\bnu)\pi(\bw)\prod_{i=1}^n \sum_{h=1}^H\nu_h\, \tilde \p(y_i \mid \btheta^h)\mbox{d}\btheta\mbox{d}\bnu < \infty
\end{equation}
Since the priors specified in \cref{prior} are proper,  it is sufficient to show that 
\begin{equation}
	\label{bound2}
\sup_{\btheta,\bnu}\, \prod_{i=1}^n \sum_{h=1}^H\nu_h\, \tilde \p(y_i \mid \btheta^h)< \infty 
\end{equation}
which can be easily proofed by expressing $\p(y_i \mid \btheta^h)$ as
\begin{equation}
	\tilde \p(y_i \mid \btheta^h) =	\prod_{{E_2 \in \mathcal{P}_2}}\exp\left\{ \sum_{\bi_{E_2}\in \I_{E_2}} \mathbb{I}(y_{i}, \bi_{E_2})\X_2 \btheta_{E_2}^h -  \kappa_2(\btheta_{E_2}^h)\right\},
\end{equation}
which is always bounded being a product of probabilities.
\end{proof}

\begin{sidewaystable}
	\caption[\textsc{scl-90} questionnaire. Subscales of interest.]{\textsc{scl-90} subscales. Subjects answer with their level of agreement with numbers ranging from $0$ (``Not at all'') to $4$ (``Extremely'').}
	\label{app:sclQ}
\centering
\begin{tabular}{lllllll}
	\textsc{id} &  &\textsc{sub} & 	\textsc{id} & & \textsc{sub} \\
\toprule
 2.  & Nervousness or shakiness inside                               & \textsc{(anx) } & 45.     & Having to check and double-check what you do       & \textsc{(oc)} \\
 3.  & Unwanted thoughts, words, or ideas that won’t leave your mind & \textsc{(oc)  } & 46.     & Difficulty making decisions                        & \textsc{(oc)} \\
 5.  & Loss of sexual interest or pleasure                           & \textsc{(dep) } & 51.     & Your mind going blank                              & \textsc{(oc)} \\
 9.  & Trouble remembering things                                    & \textsc{(oc)  } & 55.     & Trouble concentrating                              & \textsc{(oc)} \\
 10. & Worried about sloppiness or carelessness                      & \textsc{(oc)  } & 63.     & Having urges to beat, injure, or harm someone      & \textsc{(hos)} \\
 11. & Feeling easily annoyed or irritated                           & \textsc{(hos) } & 65.     & Having to repeat the same actions such as    --      & \textsc{(oc)} \\
 14. & Feeling low in energy or slowed down                          & \textsc{(dep) } &  & touching, counting, washing                        & $\cdot$ \\
 15. & Thoughts of ending your life                                  & \textsc{(dep) } & 67.     & Having urges to break or smash things              & \textsc{(hos)} \\
 17. & Trembling                                                     & \textsc{(anx) } & 71.     & Feeling everything is an effort                    & \textsc{(dep)} \\
 20. & Crying easily                                                 & \textsc{(dep) } & 72.     & Spells of terror or panic                          & \textsc{(anx)} \\
 22. & Feeling of being trapped or caught                            & \textsc{(dep) } & 74.     & Getting into frequent arguments                    & \textsc{(hos)} \\
 23. & Suddenly scared for no reason                                 & \textsc{(anx) } & 78.     & Feeling so restless you couldn’t sit still         & \textsc{(anx)} \\
 26. & Blaming yourself for things                                   & \textsc{(dep) } & 79.     & Feelings of worthlessness                          & \textsc{(dep)} \\
 28. & Feeling blocked in getting things done                        & \textsc{(oc)  } & 80.     & Feeling that familiar things are strange or unreal & \textsc{(anx)} \\
 29. & Feeling lonely                                                & \textsc{(dep) } & 86.     & Feeling pushed to get things done                  & \textsc{(anx)} \\
 30. & Feeling blue                                                  & \textsc{(dep) } & \\
 31. & Worrying too much about things                                & \textsc{(dep) } & \\
 32. & Feeling no interest in things                                 & \textsc{(dep) } & \\
 33. & Feeling fearful                                               & \textsc{(anx) } & \\
 38. & Having to do things very slowly to insure correctness         & \textsc{(oc)  } & \\
 39. & Heart pounding or racing                                      & \textsc{(anx) } & \\
\bottomrule
\end{tabular}
\end{sidewaystable}
\begin{sidewaystable}[ht]
	\caption[\textsc{iri-28} questionnaire.]{\textsc{iri-28} questionnaire. Subjects answer with their level of agreement with letters ranging from A (``Does not describe me'') to E (``Describes me very well'').}
	\label{app:iriQ}
\centering
\begin{tabular}{lll}
	\textsc{id} & & \textsc{sub} \\
\toprule
1.& I daydream and fantasize, with some regularity, about things that might happen to me.  &(\textsc{fs})\\
2.& I often have tender, concerned feelings for people less fortunate than me. &(\textsc{ec})\\
3.& I sometimes find it difficult to see things from the "other guy's" point of view. &(\textsc{pt}) \\
4.& Sometimes I don't feel very sorry for other people when they are having problems. &(\textsc{ec}) \\
5.& I really get involved with the feelings of the characters in a novel. &(\textsc{fs})\\
7.& I am usually objective when I watch a movie or play, and I don't often get completely caught up in it. &(\textsc{fs}) \\
8.& I try to look at everybody's side of a disagreement before I make a decision. &(\textsc{pt})\\
9.& When I see someone being taken advantage of, I feel kind of protective towards them.  &(\textsc{ec})\\
10.& I sometimes feel helpless when I am in the middle of a very emotional situation. &(\textsc{pd})\\
11.& I sometimes try to understand my friends better by imagining how things look from their perspective. &(\textsc{pt}) \\
12.& Becoming extremely involved in a good book or movie is somewhat rare for me. &(\textsc{fs}) \\
13.& When I see someone get hurt, I tend to remain calm. &(\textsc{pd}) \\
14.& Other people's misfortunes do not usually disturb me a great deal. &(\textsc{ec}) \\
15.& If I'm sure I'm right about something, I don't waste much time listening to other people's arguments. &(\textsc{pt}) \\
16.& After seeing a play or movie, I have felt as though I were one of the characters. &(\textsc{fs})\\
17.& Being in a tense emotional situation scares me. &(\textsc{pd})\\
18.& When I see someone being treated unfairly, I sometimes don't feel very much pity for them. &(\textsc{ec}) \\
19.& I am usually pretty effective in dealing with emergencies. &(\textsc{pd}) \\
21.& I believe that there are two sides to every question and try to look at them both. &(\textsc{pt})\\
23.& When I watch a good movie, I can very easily put myself in the place of a leading character. &(\textsc{fs})\\
25.& When I'm upset at someone, I usually try to "put myself in his shoes" for a while. &(\textsc{pt})\\
26.& When I am reading an interesting story or novel, I imagine how I would feel if the events in the story were happening to me. &(\textsc{fs})\\
28.& Before criticizing somebody, I try to imagine how I would feel if I were in their place.  &(\textsc{pt}) \\
\bottomrule
\end{tabular}
\end{sidewaystable}

\bibliographystyle{apalike}
\bibliography{ref.bib}
\end{document}